\documentclass[sigconf]{acmart}
\AtBeginDocument{%
  }

\usepackage[framemethod=tikz]{mdframed}
\usepackage{enumitem}
\usepackage{listings}
\usepackage{graphicx}
\usepackage{subcaption}
\usepackage{multirow}
\usepackage{amsmath}
\usepackage{threeparttable}
\usepackage{makecell}
\usepackage{pgfplots}
\pgfplotsset{compat=1.18}
\usetikzlibrary{positioning}

\newmdenv[
  linecolor=blue!70!white,
  linewidth=1pt,
  roundcorner=4pt,
  backgroundcolor=white,
  skipabove=\baselineskip,
  skipbelow=\baselineskip,
  frametitlefont=\bfseries,
  frametitlebackgroundcolor=blue!10!white
]{redbox}

\setcopyright{acmlicensed}
\copyrightyear{2018}
\acmYear{2018}
\acmDOI{XXXXXXX.XXXXXXX}
\acmConference[Conference acronym 'XX]{Make sure to enter the correct
  conference title from your rights confirmation email}{June 03--05,
  2018}{Woodstock, NY}
\acmISBN{978-1-4503-XXXX-X/2018/06}

\begin{document}

\title{Route What Remains: A Meta-Modal Agent for Missing-Modality Candidate Reranking in Recommender Systems}

\author{Jinze Wang$^\dag$}
\affiliation{%
    \institution{GridAIX Pty Ltd}
  \city{Melbourne}
  \country{Australia}}
  \email{jinzewang@gridaix.com.au}

\author{Yangchen Zeng$^\dag$}
\affiliation{%
  \institution{School of Computer Science and Technology}
  \institution{Southeast University}
  \city{Shanghai}
  \country{China}}

  \author{Tiehua Zhang*}
\affiliation{%
  \institution{School of Computer Science and Technology}
  \institution{Tongji University}
  \city{Shanghai}
  \country{China}}
  \email{tiehuaz@tongji.edu.cn}

\author{Lu Zhang}
\affiliation{%
  \institution{School of Cybersecurity}
  \institution{Chengdu University of Information Technology}
  \city{Chengdu}
  \country{China}}

\author{Yuze Liu}
\affiliation{%
  \institution{School of Engineering}
  \institution{Swinburne University of Technology}
  \city{Melbourne}
  \country{Australia}}

\author{Zhishu Shen}
\affiliation{%
  \institution{School of Artificial Intelligence}
  \institution{Wuhan University of Technology}
  \city{Wuhan}
  \country{China}
}

\author{Zhu Sun}
\affiliation{%
 \institution{Singapore University of Technology and Design}
 \country{Singapore}}

\renewcommand{\shortauthors}{Jinze Wang et al.}

\begin{abstract}
  Missing-modality recommenders usually reconstruct absent representations, although the observed evidence may not determine the missing content. We formulate candidate reranking as budgeted sequential evidence acquisition. A policy queries text, image, and interaction-graph tools, incorporates \texttt{Null} returns into its observation history, and sparsely rescores a retrieved candidate pool. Our \textbf{Meta-Modal Agent} (MMA) uses PPO to optimize terminal NDCG and tool cost without explicit access to the route-availability mask or target identity. When only one evidence route is available, MMA-Auto improves NDCG@10 by $10.0$\% over the strongest completion baseline and by $9.5$\% over a fixed router with the same Llama scorer. It obtains the highest result in all nine reported combinations of dataset and available route against these comparators. MMA-Auto also reduces failed calls by 17.8 percentage points and uses 1.1 fewer turns than the fixed router. On the fixed candidate pools produced by full-catalog retrieval, MMA-Auto improves NDCG@10 by $19.7$\%. These results associate adaptive evidence routing with improved reranking under severe, constructed missingness. The code is available at: https://anonymous.4open.science/r/WSDM2027-MMA-C381.
\end{abstract}



\maketitle

\section{Introduction}

Modern recommender systems rely on complementary evidence: interaction graphs represent collaborative preference, text describes item semantics, and images capture attributes absent from structured metadata \cite{liu2024multimodal,wang2025hyperman,dong2026language}. However, evidence incompleteness is often most acute in challenging recommendation settings. New users lack behavioral histories, newly listed items may contain images without descriptions, and privacy-preserving deployments may suppress user attributes~\cite{liu2024multimodal,dong12026towards}. Missing modalities therefore constitute a central cold-start condition rather than a peripheral data-processing issue~\cite{pan2022multimodal}.

Most missing-modality recommenders represent the absent signal as a latent variable to be reconstructed or compensated for. They learn predictors robust to incomplete inputs, propagate features across graphs, or generate missing embeddings from available modalities \cite{wang2018lrmm,wu2024deep,kim2025disentangling}. These approaches assume that the observed evidence sufficiently constrains the missing representation. Under severe missingness, however, a short title may support several incompatible behavioral profiles, and a noisy product image may not determine textual attributes. Consequently, a reconstruction may be plausible in representation space while introducing evidence that was not observed during ranking~\cite{malitesta2024we,zeng2026trialigngr}. The resulting challenge is to identify and exploit reliable observed evidence without requiring completion of every modality.

Reranking agents based on large language models (LLMs) provide an alternative formulation. An agent can select a tool, inspect the returned evidence, and revise its action sequence after a failed call instead of producing a single fused representation~\cite{dai2026omg,wang2026agent4poi,zeng2026deepinterestgr}. Generative agents have been applied to recommendation, and large-scale tool-use training enables LLMs to invoke external interfaces effectively \cite{qin2024toolllm,zeng2026deep,zeng2026trialigngr}. ReAct-style prompting further demonstrates how language models can interleave reasoning with environment actions \cite{yao2022react}. Static agent prompts, however, often assume that every tool returns a valid observation. A history lookup that returns \texttt{Null} may therefore cause repeated calls to an unavailable route or unsupported inference from absent evidence.

This paper formulates missingness as a \emph{sequential evidence-routing} problem within candidate-pool reranking. Because route availability is hidden, a reranker must identify usable routes, determine whether an additional query justifies its cost, and aggregate acquired evidence into candidate scores. We model this interaction as a partially observable Markov decision process (POMDP) and propose the \textbf{Meta-Modal Agent} (MMA), an LLM policy that routes among text, image, and interaction-graph tools. MMA records evidence and \texttt{Null} returns in its interaction history and conditions subsequent acquisition and stopping decisions on that history.

Sequential routing is tightly coupled with reranking. Each tool call returns candidate evidence or reveals that a route is unavailable, while consuming a limited turn budget. Because MMA assigns new scores only to candidates for which it has acquired evidence, route and candidate selection determine how sparse agent scores interact with first-stage scores over the complete pool. The policy must therefore decide which candidate and route to inspect, interpret \texttt{Null} as information about the episode, and stop when further acquisition is unlikely to improve the ranking enough to justify its cost. This coupling motivates optimization over complete routing and ranking trajectories rather than independent prediction of missing representations or isolated tool selection.

We train the policy with proximal policy optimization (PPO) over episodes constructed from a first-stage candidate pool and a hidden route-availability mask. The policy generates a trajectory of structured tool actions and terminates by scoring the inspected candidates. Its reward combines terminal normalized discounted cumulative gain (NDCG) with action costs, favoring trajectories that acquire discriminative evidence while limiting failed and unnecessary calls. Training uniformly covers seven route-availability families. At inference, the learned policy adapts through its observation history without parameter updates. We evaluate the resulting automated policy, MMA-Auto, with two complementary candidate-pool protocols. Target-positive pools use route-specific hard negatives to isolate ordering quality, whereas unmodified pools from full-catalog retrieval evaluate downstream reranking after first-stage recall. In both protocols, all methods receive the same candidate pool, and MMA observes neither the route-availability mask nor the target identity.

Our contributions are:
\begin{itemize}
    \item We formulate missing-modality candidate reranking as a POMDP in which evidence and \texttt{Null} returns update an LLM policy's acquisition trajectory. This formulation treats missingness as decision-relevant feedback rather than a representation that must always be reconstructed.
    \item We develop a budgeted sparse reranker that allocates evidence calls across routes and candidates, terminates with scores for the inspected subset, and uses PPO to optimize terminal ranking quality and acquisition cost across seven route-availability families.
    \item We design complementary target-positive and full-catalog protocols that evaluate controlled candidate ordering and downstream reranking, respectively. Both use shared candidate pools without revealing the target identity or route-availability mask to MMA.
    \item In all nine reported combinations of dataset and available route, MMA-Auto achieves the highest NDCG@10 among DGMRec, GRE-MC, and a fixed router with the same scorer. It improves average NDCG@10 by $10.0$\% over the strongest completion baseline and by $9.5$\% over the fixed router. On fixed pools produced by full-catalog retrieval, it improves NDCG@10 by $19.7$\%.
\end{itemize}

\begin{figure*}[t]
\centering
\includegraphics[width=\textwidth]{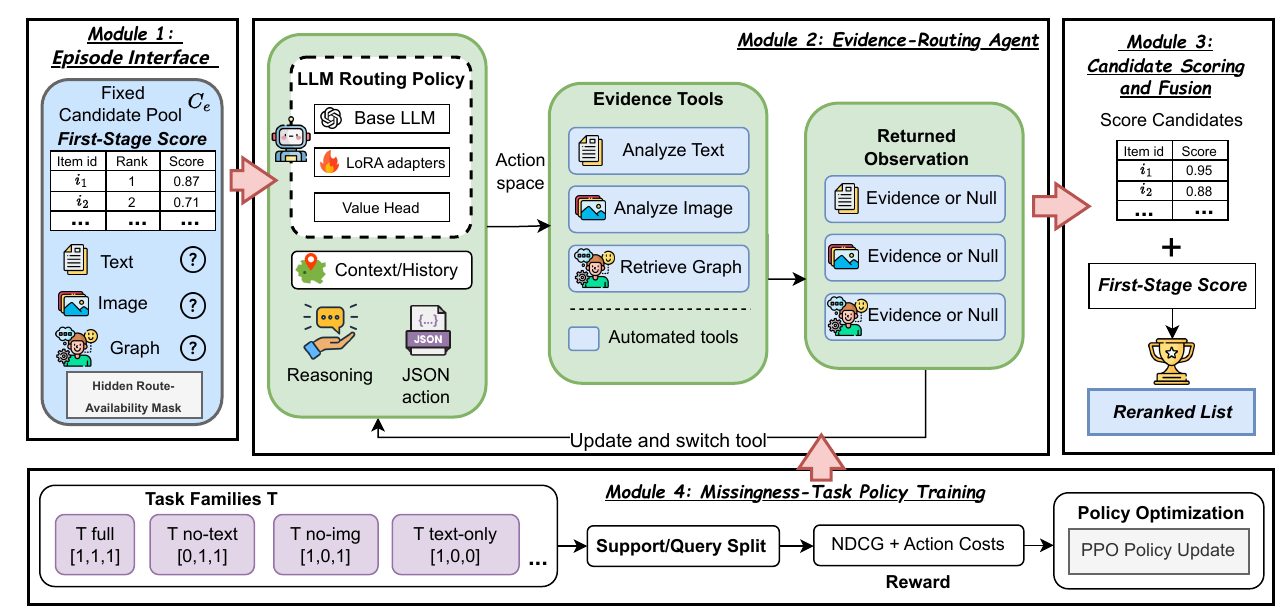}
\caption{Overview of MMA. The policy receives a fixed candidate pool without the route-availability mask or target identity. It sequentially queries text, image, and interaction-graph tools, appends evidence or \texttt{Null} returns to its history, and terminates with sparse candidate scores that are fused with first-stage scores. PPO optimizes terminal NDCG and action costs across seven route-availability families.}
\Description{Diagram showing the Meta-Modal Agent flow from a shared candidate pool through evidence-tool routing, candidate scoring, and fusion reranking under missing modalities.}
\label{fig:teaser}
\end{figure*}

\section{Related Work}

\noindent\textbf{Multimodal Recommendation under Missing Modalities.} Multimodal recommender systems use text, images, audio, and interaction graphs to improve preference modeling \cite{liu2024multimodal}. Research on missing modalities spans recommendation and broader multimodal learning \cite{wu2024deep}. Representation-based approaches address cold-start missingness through modality-specific factorization \cite{wang2018lrmm}, disentangled feature reconstruction \cite{kim2025disentangling}, graph-retrieval-enhanced completion \cite{li2026robust}, modality-aware multi-intention learning \cite{yang2024multimodal}, and multimodal graph neural networks \cite{wang2025hyperman}. UniSRec addresses sparse item representations through universal sequence pre-training \cite{hou2022towards}. These methods learn representations for retrieval or ranking, often by completing or compensating for absent inputs. MMA instead acquires observed evidence from available routes before reranking a fixed candidate pool.

\noindent\textbf{Agentic and Tool-Based Recommendation.} LLM agents maintain interaction context, call external tools, and revise action sequences after receiving observations. ReAct interleaves reasoning traces with environment actions \cite{yao2022react}, while large-scale tool-use training improves API invocation \cite{qin2024toolllm}. Recommendation agents include OMG-Agent and Agent4POI \cite{dai2026omg,wang2026agent4poi}, generative recommendation agents \cite{zhang2024generative,wang2025we}, and AgentCF's simulated user--item agents \cite{zhang2024agentcf}. InstructRec instead adapts LLMs to natural-language recommendation instructions across item domains \cite{zhang2026recommendation}. MMA shares the agentic interface but focuses on a distinct decision problem: discovering unavailable routes through tool feedback, allocating a bounded acquisition budget, and producing sparse scores over a fixed candidate pool.

\noindent\textbf{Sequential Evidence Acquisition.} RL has been applied to sequential recommendation \cite{xin2020self} and conversational recommendation, where an agent asks attribute questions before recommending \cite{lei2020estimation}. RLMRec incorporates RL signals into multimodal recommendation to align item representations with user preferences \cite{ren2024representation}. MMA acquires evidence from system tools rather than soliciting user clarification, and it optimizes terminal reranking quality jointly with acquisition cost rather than representation alignment. Multi-task and curriculum training can improve sample efficiency and generalization \cite{wang2023meta}; MMA balances seven route-availability families so that severe missingness patterns receive comparable training exposure. In this work, the term ``Meta-Modal'' refers to a shared routing policy learned across route-availability families.

\noindent\textbf{LLM-Based Recommendation.} Recent work increasingly formulates recommendation as a language task \cite{bao2023large}. LlamaRec uses a two-stage pipeline in which a retrieval model supplies candidates for LLM reranking \cite{yue2023llamarec}. We adopt this separation between first-stage retrieval and LLM-based reranking but study route discovery and cost-constrained evidence acquisition within the reranking stage. A \texttt{Null} observation enters MMA's action history, allowing later decisions to condition on unavailable evidence rather than assuming dense context.

\section{Methodology}

MMA is a candidate-pool reranker. First-stage retrieval recalls items from the catalog, after which MMA acquires evidence and rescores selected items. Figure~\ref{fig:teaser} presents four connected modules. Module 1 instantiates an episode with a candidate pool and a hidden route-availability mask. Module 2 generates a history-dependent sequence of evidence actions and observations. Module 3 terminates the sequence with candidate scores and a fused ranking. Module 4 samples masks, collects routing and ranking trajectories, evaluates ranking rewards and action costs, and updates the Module 2 policy with PPO. At test time, the learned policy executes Modules 1--3 and adapts through its observation history without parameter updates.

\subsection{Module 1: Episode Interface and Candidate Pool}
\label{sec:pomdp}

Each episode \(e\) begins with a shared candidate pool
\(C_e \subset \mathcal{I}\) of size \(B\), i.e.,
\(|C_e| = B\). For every candidate, the episode interface provides an anonymized item ID, its first-stage rank, and normalized first-stage score $s_i^{(0)}$. The initial score is derived from the first-stage retriever (DGMRec) and min--max normalized over $C_e$:
\begin{equation}
\label{eq:init_score}
s_i^{(0)} = \frac{f_i - \min_{j\in C_e} f_j}{\max_{j\in C_e} f_j - \min_{j\in C_e} f_j},
\end{equation}
where $f_i$ is the raw retriever score for item $i$. When all candidates in $C_e$ share the same raw score (a degenerate pool), we set $s_i^{(0)}=0.5$ for all $i\in C_e$ to avoid division by zero. Evidence content is accessible only through the tools in Module 2: text queries return item metadata, image queries return semantic tag summaries, and interaction-graph queries return graph-neighbor summaries. The evaluator uses the held-out target $i^*$ to construct the candidate pool, compute the reward, and evaluate the final ranking. Neither its identity nor the route-availability mask enters the policy context.

We represent route availability by a binary mask $\mathbf{m} \in \{0,1\}^{|\mathcal{M}|}$, where $\mathcal{M}=\{m_{\text{text}},m_{\text{img}},m_{\text{graph}}\}$ denotes the evidence routes. The mask remains fixed throughout each episode but varies across episodes according to the task-family distribution defined in Section~\ref{sec:curriculum}. The mask is not explicitly revealed to the agent; instead, route availability is inferred through tool interactions based on whether the environment returns valid evidence or \texttt{Null}.

The target-positive protocol includes \(i^\ast\) in \(C_e\). It retrieves the remaining \(B-1\) hard negatives using the target item's representation under the available evidence route as the anchor. Route-specific retrieval is applied to text, images, or the interaction graph. Pool construction occurs inside the evaluator, and every method receives the resulting shared pool without the target label. The first-stage retriever then assigns a raw score $f_i$ to every candidate, including the target, and ranks candidates in descending order of $f_i$. The target therefore receives neither a target-specific score nor a designated insertion rank. MMA operates over $C_e$ as a candidate scorer and reranker. Section~\ref{sec:experiments} separately reports fixed-pool evaluation after full-catalog retrieval.

\textbf{Initial context and route discovery.} The compact initial context (anonymized candidate IDs, first-stage ranks, and scores) has the same schema across episode types and contains neither evidence snippets nor an explicit mask indicator. Score distributions may provide statistical cues, but they do not certify route availability. Tool returns provide direct feedback: a non-null observation identifies a usable route, whereas a \texttt{Null} observation records that the queried route is unavailable.

\subsection{Module 2: Evidence-Routing Agent}

We formulate each episode as a partially observable Markov decision process (POMDP) $\langle \mathcal{S}, \mathcal{A}, \mathcal{P}, \Omega, \mathcal{O}, \mathcal{R}, \gamma \rangle$. The latent state $s \in \mathcal{S}$ contains the user preference distribution over $\mathcal{I}$, candidate evidence, and the fixed observability mask $\mathbf{m}$. The observation function $\Omega(o_t\mid s,a_t)$ returns evidence when the queried route is available and \texttt{Null} otherwise. The agent does not maintain an explicit parametric belief state; instead, the LLM uses the accumulated interaction history
\[
H_t = (z_1,a_1,o_1,\ldots,z_t,a_t,o_t),
\]
where $z_t$ is the reasoning trace, $a_t$ is the structured action, and $o_t$ is the returned observation. The environment state is static within an episode, so $\mathcal{P}(s' \mid s,a_t)=\mathbf{1}[s'=s]$; the decision state nevertheless changes because $H_t$ records which paths succeeded, which returned \texttt{Null}, and what candidate evidence has already been collected. This history lets the policy infer availability and compare the expected value of another call against terminating with the current evidence.

Although the policy represents this state implicitly in its context, the information supplied by a tool return has a simple belief-space interpretation. Let $b_t(\mathbf{m})=p(\mathbf{m}\mid H_t)$ denote a conceptual posterior over route-availability masks. After action $a_t$ and observation $o_t$,
\begin{equation}
b_t(\mathbf{m}) \propto \Omega(o_t\mid \mathbf{m},a_t)b_{t-1}(\mathbf{m}).
\label{eq:belief}
\end{equation}
With deterministic availability, a \texttt{Null} return removes every mask in which the queried route is present; a successful return removes every mask in which it is absent. Because the initial context does not explicitly specify the mask, each return contracts the feasible mask set. MMA learns how to condition subsequent acquisition and stopping decisions on this feedback and the evidence already collected.

At turn $t$, the LLM policy first generates a reasoning trace, followed by a single-line JSON action:
\begin{equation}
z_t \sim \pi_\theta^z(z \mid H_{t-1}), \qquad a_t \sim \pi_\theta^a(a \mid H_{t-1}, z_t).
\label{eq:policy}
\end{equation}
The action space is $\mathcal{A}=\mathcal{N}\times\mathcal{Q}$, where $\mathcal{N}$ contains the three evidence tools and the terminal \texttt{Score\_Candidates} action, and $\mathcal{Q}$ is the corresponding JSON argument string. A routing decision specifies which route to inspect and, for item-level tools, which candidate to query. Invalid JSON or an unknown tool name incurs the cost in Eq.~\eqref{eq:action_cost}.

We count every generated structured action as one turn. Thus, an evidence request, a failed or malformed request, and the terminal \texttt{Score\_Candidates} action each consume one unit of the budget. This convention aligns the optimization horizon, the reported turn count, and the inference-cost calculation. Termination before $T$ is an explicit policy decision rather than an external early-exit heuristic.

\textbf{Evidence tools.} MMA accesses three automated tools:
\begin{itemize}
    \item \texttt{Analyze\_Text(item\_id)} returns normalized title, category, and truncated description fields; it returns \texttt{Null} if $m_{\text{text}}=0$.
    \item \texttt{Analyze\_Image(item\_id)} runs a frozen vision-language encoder and returns zero-shot semantic tags; it returns \texttt{Null} if $m_{\text{img}}=0$.
    \item \texttt{Retrieve\_Graph()} is bound to the episode's user context. It performs ANN search over a precomputed item--item collaborative graph. Each returned candidate-pool entry contains a candidate ID, title, and graph score; the call returns \texttt{Null} if $m_{\text{graph}}=0$.
\end{itemize}

The tool interface incorporates absence into the decision process. Text and image observations describe a queried candidate. The episode-bound graph call returns candidate-linked collaborative evidence restricted to $C_e$; its titles are associated metadata rather than a separate text query. A successful call adds evidence to $H_t$, whereas a \texttt{Null} call marks the queried route as unavailable. The policy can then select another route or terminate with the evidence already acquired.

\subsection{Module 3: Candidate Scoring and Fusion}
\label{sec:scoring}

Instead of retrieving an item or emitting a single recommendation, MMA terminates with \texttt{Score\_Candidates}, a non-empty JSON map from evidence-eligible candidate IDs to finite relevance scores $s_i^{\mathrm{MMA}}$. Let $E_e\subseteq C_e$ contain the IDs queried successfully through the text or image tools together with the candidate IDs returned by the graph tool. A valid terminal map contains IDs $Q_e$ satisfying $\varnothing\neq Q_e\subseteq E_e\subseteq C_e$. An empty map, an ID outside $E_e$, a duplicate ID, or a non-numeric or non-finite score (including \texttt{NaN} or \texttt{Inf}) is an invalid terminal output. Candidates outside $Q_e$ retain their initial scores. This contract preserves the candidate set and prevents the terminal action from assigning evidence-based scores to items for which no candidate-level evidence was acquired.

MMA scores are min--max normalized over $Q_e$ before fusion. Let $\sigma_{\max}=\max_{i\in Q_e}s_i^{\mathrm{MMA}}$ and $\sigma_{\min}=\min_{i\in Q_e}s_i^{\mathrm{MMA}}$. The normalized MMA score is
\[
\tilde{s}_i^{\mathrm{MMA}} =
\begin{cases}
\dfrac{s_i^{\mathrm{MMA}}-\sigma_{\min}}{\sigma_{\max}-\sigma_{\min}} & \text{if } \sigma_{\max}>\sigma_{\min},\\
0.5 & \text{if } |Q_e|=1 \text{ or } \sigma_{\max}=\sigma_{\min}.
\end{cases}
\]
The fused score for queried candidates is
\begin{equation}
\label{eq:fusion}
    \hat{s}_i = \alpha\, s_i^{(0)} + (1-\alpha)\, \tilde{s}_i^{\mathrm{MMA}}, \quad i\in Q_e.
\end{equation}

The fusion weight $\alpha$ is selected on the validation split and then fixed for test evaluation. Equation~\eqref{eq:fusion} is applied only to $Q_e$; candidates in $C_e\setminus Q_e$ keep their first-stage scores. The complete ranking is ordered by fused score, with the first-stage rank and then candidate ID used as deterministic tie-breakers. The normalization range therefore depends only on scores emitted by the agent. An invalid terminal output immediately ends the episode and invokes the first-stage fallback defined below. HR@$K$ and NDCG@$K$ are computed from the resulting complete ordering of $C_e$.

\textbf{Sparse scoring under a turn budget.} Because the turn budget $T$ is smaller than the candidate-pool size $B$, direct evidence retrieval for every candidate is unavailable. MMA therefore operates as a \emph{sparse reranker}. It queries evidence for a small set of high-priority candidates and assigns $s_i^{\mathrm{MMA}}$ only to items included in $Q_e$. Unqueried candidates retain $\hat{s}_i = s_i^{(0)}$, which is equivalent to setting $\alpha=1$ for those items. The first-stage score distribution guides candidate selection, while the PPO objective rewards acquisitions that improve the final ranking. A bounded number of evidence observations can thus modify the head of a larger candidate pool without assigning evidence-based scores to uninspected items.

\Needspace{0.24\textheight}
\begin{redbox}[frametitle={Budgeted routing and reranking}]
\small
\textbf{Input:} candidate-score pairs $\{(i,s_i^{(0)})\}_{i\in C_e}$; budget $T$.\\
\textbf{1.} Initialize $H_0$ with anonymized IDs, ranks, and scores.\\
\textbf{2.} For $t=1,\ldots,T$, sample a structured action from Eq.~\eqref{eq:policy}.\\
\textbf{3.} For an evidence action, append its evidence or \texttt{Null} return to $H_t$.\\
\textbf{4.} For a valid \texttt{Score\_Candidates}, apply Eq.~\eqref{eq:fusion} and stop.\\
\textbf{5.} For an invalid terminal output, stop immediately and use the first-stage fallback.\\
\textbf{6.} At context truncation or budget exhaustion without a valid terminal action, use the same fallback.
\end{redbox}

The online routing cost is at most $T$ LLM decisions and at most $T-1$ evidence calls because one action is required for normal termination. This cost does not scale with catalog size after retrieval; the ranking step is $O(B\log B)$. Exhaustive multimodal inspection would instead require $O(B|\mathcal{M}|)$ evidence calls. The design therefore decouples catalog-scale recall from a bounded evidence-acquisition budget.

\subsection{Module 4: Missingness-Task Policy Training}
\label{sec:curriculum}

If rollouts follow the naturally pooled data distribution, masks with abundant evidence dominate and teach the agent to prefer tools that usually succeed. Such a policy can collapse when only one evidence route remains available. We therefore organize rollouts into the seven missingness-task families in Table~\ref{tab:tasks}, each defined by a fixed mask distribution $P_k(\mathbf{m})$. The families span full observation, three leave-one-modality-out (LOMO) conditions, and three severe OOMA conditions. The interaction-graph route provides collaborative scores together with neighbor-title metadata. Uniform task sampling gives each availability pattern equal influence on the policy gradient and makes per-pattern routing behavior directly measurable.

\begin{table}[t!]
\centering
\caption{Route-availability task families.}
\label{tab:tasks}
\small
\begin{tabular}{llccc}
\toprule
\textbf{Family} & \textbf{Condition} & $m_{\text{text}}$ & $m_{\text{img}}$ & $m_{\text{graph}}$ \\
\midrule
$\mathcal{T}_{\text{full}}$       & All routes available & 1 & 1 & 1 \\
$\mathcal{T}_{\text{no-text}}$    & Text missing (LOMO)       & 0 & 1 & 1 \\
$\mathcal{T}_{\text{no-img}}$     & Image missing (LOMO)      & 1 & 0 & 1 \\
$\mathcal{T}_{\text{cold-user}}$  & Interaction graph missing (LOMO) & 1 & 1 & 0 \\
$\mathcal{T}_{\text{text-only}}$  & Text only (OOMA)          & 1 & 0 & 0 \\
$\mathcal{T}_{\text{img-only}}$   & Image only (OOMA-Visual)  & 0 & 1 & 0 \\
$\mathcal{T}_{\text{graph}}$       & Graph-route-only (OOMA-Graph) & 0 & 0 & 1 \\
\bottomrule
\end{tabular}
\end{table}

Each family is divided into support episodes $\mathcal{D}^s_k$ and disjoint query episodes $\mathcal{D}^q_k$. Support episodes provide policy updates, whereas query episodes measure transfer to unseen episodes with the same availability structure. Training rollouts sample task families uniformly,
\[
p(\mathcal{T})=\mathrm{Uniform}(\{\mathcal{T}_k\}_{k=1}^7),
\]
so severe OOMA cases receive the same training exposure as easier fully observed or single-missing-route cases.

\textbf{Reward.} The termination time $\tau$ is set by the earliest terminal event. Such an event is a valid \texttt{Score\_Candidates} action, an invalid terminal output, context-limit truncation, or exhaustion of the $T$-turn budget. The environment assigns an action cost
\begin{equation}
\label{eq:action_cost}
c_t =
\begin{cases}
\lambda_{\text{tool}} & \text{valid evidence action} \\
\lambda_{\text{invalid}} & \text{invalid or repeated failed action} \\
0 & \text{valid terminal action}.
\end{cases}
\end{equation}
The per-turn reward is then
\begin{equation}
\label{eq:reward}
\mathcal{R}_t=c_t+\mathbf{1}[t=\tau]\,\mathrm{NDCG}@K(\hat{\mathbf{r}}_{\tau},i^*).
\end{equation}

An invalid action includes malformed JSON or an unknown tool. A repeated failed action queries a route already observed to return \texttt{Null}, and an invalid terminal output follows the validation rules in Section~\ref{sec:scoring}. For a valid terminal action, $\hat{\mathbf{r}}_{\tau}$ is the ranking induced by Eq.~\eqref{eq:fusion}. An invalid terminal output ends the episode immediately, retains its invalid-action cost, and uses the unchanged first-stage ranking for terminal NDCG. Context truncation and budget exhaustion without a valid terminal action use the same fallback. If the action at turn $T$ is an evidence or invalid action, its cost is retained before terminal NDCG is evaluated; context truncation without an emitted action adds no action cost. Equation~\eqref{eq:reward} connects ranking to routing: terminal NDCG rewards trajectories that rank the target highly, while accumulated action costs favor shorter valid paths. The ordering $|\lambda_{\mathrm{tool}}|<|\lambda_{\mathrm{invalid}}|$ makes a valid lookup cheaper than a malformed or repeated failed action. Exact values are reported in Appendix~\ref{app:implementation}.

\textbf{Policy optimization.} Within an episode, MMA adapts only through context $H_t$; no gradient update occurs at test time. During training, we optimize expected reward over support episodes:
\begin{equation}
\label{eq:objective}
    \max_\theta\; \mathcal{J}(\theta) =
    \mathbb{E}_{\mathcal{T}_k \sim p(\mathcal{T})}
    \mathbb{E}_{e \sim \mathcal{D}^s_k}
    \left[ \sum_{t=1}^{\tau} \gamma^t\, \mathcal{R}(i^*_e,a_t,H_{t-1}) \right].
\end{equation}

The optimization variables are the LoRA policy parameters $\theta$ and value-head parameters $\phi$; the tool backends and base LLM remain fixed. We use PPO because the policy emits discrete structured actions and receives a sparse, non-differentiable ranking reward. Policy gradients optimize this objective directly, while PPO clipping limits updates dominated by a small number of high-reward trajectories. Generalized advantage estimation assigns temporal credit by propagating terminal NDCG and intermediate costs from Eq.~\eqref{eq:reward} to each history--action pair $(H_{t-1},(z_t,a_t))$:
\begin{equation}
\hat{A}_t = \sum_{l=0}^{\tau-t}(\gamma\lambda)^l\delta_{t+l}, \quad
\delta_t = r_t+\gamma V_\phi(H_{t+1})-V_\phi(H_t),
\label{eq:gae}
\end{equation}
and the clipped surrogate objective
\begin{equation}
\label{eq:ppo}
\mathcal{L}^{\text{CLIP}}(\theta)=
\hat{\mathbb{E}}_t
\left[
\min\left(
\rho_t(\theta)\hat{A}_t,
\mathrm{clip}(\rho_t(\theta),1-\epsilon,1+\epsilon)\hat{A}_t
\right)
\right],
\end{equation}
where $\epsilon$ is the PPO clipping threshold and
\[
\rho_t(\theta)=
\frac{\pi_\theta(z_t,a_t\mid H_{t-1})}
{\pi_{\theta_{\mathrm{old}}}(z_t,a_t\mid H_{t-1})}.
\]

In the POMDP formulation, $H_{t-1}$ is the policy input induced by the latent state and previous observations, $(z_t,a_t)$ is the reasoning and action pair sampled in Eq.~\eqref{eq:policy}, and $r_t=\mathcal{R}_t$ is the step or terminal reward from Eq.~\eqref{eq:reward}. The value function $V_\phi(H_t)$ estimates the remaining return until episode termination, with terminal bootstrap $V_\phi(H_{\tau+1})=0$. PPO consequently optimizes the routing behavior defined in Module 2.

\textbf{Training controls.} The zero-shot ReAct baseline receives the tool schema without supervised fine-tuning (SFT) or RL training. Its comparison with MMA-SFT and MMA-Auto distinguishes prompt-based tool access, supervised action formatting, and policy optimization. Appendix Table~\ref{tab:full_baselines} reports the corresponding results.

\begin{table}[t]
\centering
\caption{Dataset statistics after chronological splitting.}
\label{tab:datasets}
\small
\begin{tabular}{lrrrr}
\toprule
\textbf{Dataset} & \textbf{\#Users} & \textbf{\#Items} & \textbf{\#Inter.} & \textbf{Sparsity} \\
\midrule
Baby   & 19,445 & 7,050  & 160,792 & 99.88\% \\
Sports & 35,598 & 18,357 & 296,337 & 99.95\% \\
Yelp   & 30,887 & 20,033 & 368,340 & 99.94\% \\
\bottomrule
\end{tabular}
\end{table}

\section{Experiments}
\label{sec:experiments}

We organize the evaluation around four questions. MMA-Auto operates in a \emph{mask-hidden} setting: at inference, neither the route-availability mask nor the held-out target identity is explicitly supplied to the policy. Initial scores may carry statistical cues, while tool returns provide direct route-availability feedback.
\begin{itemize}
    \item \textbf{RQ1:} Does mask-hidden MMA-Auto improve OOMA reranking over strong static completion baselines?
    \item \textbf{RQ2:} Does the OOMA advantage hold for text-only, image-only, and graph-route-only episodes?
    \item \textbf{RQ3:} Does MMA-Auto remain effective when a full-catalog first-stage retriever supplies a fixed pool for reranking?
    \item \textbf{RQ4:} How does MMA compare with controlled fixed routers in ranking quality, routing behavior, and inference cost?
\end{itemize}
Table~\ref{tab:ooma_breakdown} reports per-route target-positive results as mean $\pm$ standard deviation over five seeds. LLM-Fixed-Router uses the same scorer as MMA-Auto with a fixed routing order.
\begin{table*}[t]
\centering
\caption{Per-route OOMA NDCG@10 under target-positive evaluation.}
\label{tab:ooma_breakdown}
\small
\begin{tabular}{llcccc}
\toprule
Dataset & Method & Text-only & Image-only & Graph-route-only & Avg. OOMA \\
\midrule
Baby & DGMRec & $0.2016_{\pm.0092}$ & $0.1620_{\pm.0080}$ & $0.1908_{\pm.0088}$ & $0.1848_{\pm.0088}$ \\
Baby & GRE-MC & $0.1940_{\pm.0096}$ & $0.1580_{\pm.0084}$ & $0.1880_{\pm.0092}$ & $0.1800_{\pm.0096}$ \\
Baby & LLM-Fixed-Router & $0.2014_{\pm.0090}$ & $0.1628_{\pm.0082}$ & $0.1896_{\pm.0088}$ & $0.1846_{\pm.0092}$ \\
Baby & MMA-Auto & $\mathbf{0.2216_{\pm.0124}}$ & $\mathbf{0.1772_{\pm.0110}}$ & $\mathbf{0.2108_{\pm.0120}}$ & $\mathbf{0.2032_{\pm.0124}}$ \\
\midrule
Sports & DGMRec & $0.1820_{\pm.0088}$ & $0.1440_{\pm.0076}$ & $0.1732_{\pm.0084}$ & $0.1664_{\pm.0080}$ \\
Sports & GRE-MC & $0.1748_{\pm.0092}$ & $0.1384_{\pm.0080}$ & $0.1668_{\pm.0088}$ & $0.1600_{\pm.0096}$ \\
Sports & LLM-Fixed-Router & $0.1836_{\pm.0084}$ & $0.1452_{\pm.0078}$ & $0.1716_{\pm.0086}$ & $0.1668_{\pm.0082}$ \\
Sports & MMA-Auto & $\mathbf{0.1996_{\pm.0112}}$ & $\mathbf{0.1610_{\pm.0100}}$ & $\mathbf{0.1884_{\pm.0108}}$ & $\mathbf{0.1830_{\pm.0110}}$ \\
\midrule
Yelp & DGMRec & $0.1560_{\pm.0080}$ & $0.1240_{\pm.0068}$ & $0.1472_{\pm.0076}$ & $0.1424_{\pm.0072}$ \\
Yelp & GRE-MC & $0.1480_{\pm.0088}$ & $0.1160_{\pm.0072}$ & $0.1392_{\pm.0080}$ & $0.1344_{\pm.0096}$ \\
Yelp & LLM-Fixed-Router & $0.1584_{\pm.0076}$ & $0.1258_{\pm.0070}$ & $0.1481_{\pm.0074}$ & $0.1441_{\pm.0078}$ \\
Yelp & MMA-Auto & $\mathbf{0.1708_{\pm.0104}}$ & $\mathbf{0.1386_{\pm.0092}}$ & $\mathbf{0.1604_{\pm.0100}}$ & $\mathbf{0.1566_{\pm.0100}}$ \\
\bottomrule
\end{tabular}
\end{table*}
\subsection{Experimental Setup}

\textbf{Datasets.} We evaluate on \textbf{Amazon-Baby} and \textbf{Amazon-Sports} \cite{mcauley2015image}, and \textbf{Yelp} (Yelp Open Dataset 2022 release). Amazon datasets provide text metadata, product images, and collaborative-filtering (CF) signals. For Yelp, text is drawn from item descriptions and review summaries; visual features are CLIP-ViT-L/14 embeddings of up to three business photos per venue; CF signals come from the user--business interaction graph. Data is split chronologically: the earliest 80\% of interactions form training, the next 10\% validation, and the most recent 10\% test. Table~\ref{tab:datasets} summarizes statistics.

\textbf{Split hierarchy.} The chronological split and the support/query split in Section~\ref{sec:curriculum} operate at different levels. We first partition interactions chronologically. Task-family episodes constructed from the training partition are then divided into 80\% support episodes for policy updates and 20\% disjoint query episodes for monitoring transfer within the same mask family. Validation interactions select fusion and stopping settings. All reported rankings use episodes generated from the held-out chronological test partition; no test interaction contributes to policy updates or model selection.

\textbf{Candidate-pool protocol.}
All methods are evaluated within the same target-positive
$B{=}100$-item candidate pool per episode.
The held-out target item is inserted into each pool. The remaining $B{-}1{=}99$ candidates are hard negatives retrieved using the target representation under the available evidence route. Retrieval uses BM25 for text, CLIP-ViT-L/14 nearest neighbors for images, and ANN search for the interaction graph. DGMRec supplies the initial retrieval scores. Pool construction occurs inside the evaluator, and every method receives the same candidates without the target label. MMA acts as a scorer and reranker over this pool and does not retrieve additional catalog items. With a turn budget $T{=}8$, MMA-Auto uses 2.1 turns per episode on average (Table~\ref{tab:det_router}). Candidates not selected for evidence acquisition retain their first-stage scores. The PPO reward therefore trains the policy to allocate acquisition to candidates for which evidence most improves the final ranking.

\textbf{Baselines.}
\textit{(1) CF:} LightGCN \cite{he2020lightgcn}, SASRec \cite{kang2018self}.
\textit{(2) Multimodal CF:} MMGCN \cite{wei2019mmgcn}, LRMM \cite{wang2018lrmm}; completely missing modalities are represented by zero-vectors under OOMA.
\textit{(3) Recommendation-specific completion baselines:} DGMRec \cite{kim2025disentangling} and GRE-MC \cite{li2026robust}.
\textit{(4) LLM agents:} ReAct zero-shot \cite{yao2022react}, AgentCF \cite{zhang2024agentcf}, Static-LLM-Ranker, MMA-SFT, and MMA-Auto. All LLM-agent variants use the same base LLM, candidate pool, context budget, and scoring protocol from Section~\ref{sec:scoring}. Static-LLM-Ranker uses the same Llama-3-8B-Instruct scorer and serialized evidence format as MMA-Auto, but scores candidates in one pass without sequential tool calling.
\textit{(5) Deterministic router controls:} RuleRouter-Fuse uses the same pool, tools, budget, and fusion rule as MMA-Auto but fixes the routing order and uses deterministic evidence scorers. LLM-Fixed-Router retains that routing order while substituting MMA's Llama scorer. Neither control observes the hidden mask; both discover missingness through \texttt{Null} returns. These controls distinguish the effects associated with semantic scoring and learned routing. Appendix Table~\ref{tab:full_baselines} reports all baselines.

\textbf{Implementation and metrics.} All agent variants share Llama-3-8B-Instruct with 8-bit quantisation, its tokenizer, the tool-schema prompts, a 2,048-token context budget, and evaluation temperature 0.0. MMA trains LoRA adapters ($r{=}16$, $\alpha{=}32$) on \texttt{q\_proj}/\texttt{v\_proj} and a two-layer value head with hidden size 256 while keeping the base weights frozen. PPO uses lr $1{\times}10^{-5}$, batch 64, clip $\epsilon{=}0.2$, GAE $\lambda{=}0.95$, $T{=}8$ turns, 500 iterations, and training temperature 0.2. The image tool uses frozen CLIP-ViT-L/14 and returns the top five zero-shot semantic tags. The fusion weight in Eq.~\eqref{eq:fusion} is selected from $\{0.1,0.2,\ldots,0.9\}$ on validation data and then fixed for test evaluation. The reward uses NDCG@10 with $\lambda_{\mathrm{tool}}=-0.02$ and $\lambda_{\mathrm{invalid}}=-0.20$. A turn is any structured action, including terminal scoring. The failed-call rate is the number of evidence-tool calls that return \texttt{Null} divided by all evidence-tool calls; it excludes the terminal scoring action. Main values are mean $\pm$ std over 5 seeds. For episode-level tests, each method's NDCG@10 is first averaged across seeds for every held-out episode; paired Wilcoxon signed-rank tests then compare matched episode scores within each dataset. Table~\ref{tab:significance} reports raw within-dataset $p$ values for the three MMA-Auto versus DGMRec comparisons; Holm--Bonferroni correction is applied only to this three-comparison family.

\subsection{Mask-Hidden OOMA Main Result (RQ1)}

Figure~\ref{fig:no_ask_main} evaluates adaptive evidence acquisition when only one evidence route remains available. Relative to DGMRec, the strongest static baseline, MMA-Auto improves average OOMA NDCG@10 by $10.0$\%, with gains on all three datasets.

\begin{figure}[t]
    \centering
    \begin{tikzpicture}
        \begin{axis}[
            ybar,
            bar width=5pt,
            width=0.85\columnwidth,
            height=0.65\columnwidth,
            enlarge x limits=0.15,
            area legend,
            legend style={at={(0.98,0.98)}, anchor=north east, legend columns=1, font=\tiny},
            ylabel={OOMA NDCG@10},
            symbolic x coords={Baby, Sports, Yelp, Avg.},
            xtick=data,
            nodes near coords,
            nodes near coords align={vertical},
            every node near coord/.append style={font=\tiny, rotate=90, anchor=west},
            ymin=0.10, ymax=0.23,
            label style={font=\scriptsize},
            tick label style={font=\scriptsize}
        ]
        \addplot[fill=blue!30] coordinates {(Baby,0.1848) (Sports,0.1664) (Yelp,0.1424) (Avg.,0.1645)};
        \addplot[fill=gray!40] coordinates {(Baby,0.1800) (Sports,0.1600) (Yelp,0.1344) (Avg.,0.1581)};
        \addplot[fill=cyan!40] coordinates {(Baby,0.1846) (Sports,0.1668) (Yelp,0.1441) (Avg.,0.1652)};
        \addplot[fill=green!60!black!60] coordinates {(Baby,0.2032) (Sports,0.1830) (Yelp,0.1566) (Avg.,0.1809)};
        \legend{DGMRec, GRE-MC, LLM-Fixed, MMA-Auto}
        \end{axis}
    \end{tikzpicture}
    \caption{Mask-hidden OOMA NDCG@10.}
    \Description{Grouped bars compare DGMRec, GRE-MC, LLM-Fixed, and MMA-Auto on Baby, Sports, Yelp, and their average. MMA-Auto has the highest NDCG at every position.}
    \label{fig:no_ask_main}
\end{figure}

MMA-Auto gains $+0.0184$, $+0.0166$, and $+0.0142$ NDCG@10 on Baby, Sports, and Yelp. \textbf{LLM-Fixed-Router} uses the same Llama-3-8B-Instruct scorer and evidence serialization. It follows the fixed route order text, graph, then image. LLM-Fixed-Router achieves $0.1652$ average NDCG@10, compared with $0.1809$ for MMA-Auto, yielding a $9.5\%$ gap under the same scorer. RuleRouter-Fuse provides a deterministic reference and reaches $0.1578$. The improvement is consistent across domains: Amazon provides structured product text and catalog images, whereas Yelp relies on review-centric text, venue photos, and a different interaction graph.

The target-positive protocol retrieves every negative from the available route, producing candidates that are plausible under the same type of evidence available to the agent. The comparison therefore measures the policy's ability to acquire candidate-specific evidence, combine it with first-stage relevance, and rank the held-out item above hard negatives. The following per-route analysis examines whether this improvement persists across the three evidence routes.

\subsection{Per-Route OOMA Breakdown (RQ2)}

Table~\ref{tab:ooma_breakdown} shows that the ranking advantage spans all three available routes. Among DGMRec, GRE-MC, LLM-Fixed-Router, and MMA-Auto, MMA-Auto obtains the highest score in all nine combinations of dataset and evidence route. The result covers semantic, visual, and collaborative evidence across all three domains.

MMA-Auto improves over LLM-Fixed-Router in all nine combinations of dataset and evidence route. On image-only episodes, the relative gaps are $+8.8$\% on Baby, $+10.9$\% on Sports, and $+10.2$\% on Yelp; the corresponding text-only gaps are $+10.0$\%, $+8.7$\%, and $+7.8$\%. On the Amazon datasets, graph-route-only scores are closer to text-only scores than to image-only scores, indicating that graph-neighbor summaries provide useful collaborative evidence. The policy conditions its routing decisions on tool feedback and the score cues available in its mask-hidden initial context.

Table~\ref{tab:full_catalog} evaluates fixed-pool full-catalog reranking under OOMA. \emph{Static} denotes the unmodified DGMRec ranking, and every reranker receives the same top-100 pool, so first-stage recall is constant. Values are mean $\pm$ standard deviation over five seeds.

\begin{table}[t]
\centering
\caption{Fixed-pool full-catalog reranking under OOMA.}
\label{tab:full_catalog}
\small
\begin{tabular}{llcc}
\toprule
Dataset & Method & HR@10 & NDCG@10 \\
\midrule
\multirow{7}{*}{Baby} 
 & Static & $0.0460_{\pm.0048}$ & $0.0212_{\pm.0024}$ \\
 & GRE-MC & $0.0442_{\pm.0046}$ & $0.0205_{\pm.0022}$ \\
 & AgentCF & $0.0384_{\pm.0040}$ & $0.0172_{\pm.0018}$ \\
 & MMA-SFT & $0.0402_{\pm.0042}$ & $0.0184_{\pm.0020}$ \\
 & RuleRouter-Fuse & $0.0448_{\pm.0044}$ & $0.0208_{\pm.0020}$ \\
 & LLM-Fixed-Router & $0.0478_{\pm.0042}$ & $0.0225_{\pm.0022}$ \\
 & \textbf{MMA-Auto} & $\mathbf{0.0544_{\pm.0044}}$ & $\mathbf{0.0254_{\pm.0022}}$ \\
\midrule
\multirow{7}{*}{Sports} 
 & Static & $0.0384_{\pm.0044}$ & $0.0180_{\pm.0020}$ \\
 & GRE-MC & $0.0362_{\pm.0042}$ & $0.0170_{\pm.0020}$ \\
 & AgentCF & $0.0318_{\pm.0038}$ & $0.0146_{\pm.0018}$ \\
 & MMA-SFT & $0.0344_{\pm.0040}$ & $0.0159_{\pm.0016}$ \\
 & RuleRouter-Fuse & $0.0378_{\pm.0042}$ & $0.0178_{\pm.0018}$ \\
 & LLM-Fixed-Router & $0.0406_{\pm.0040}$ & $0.0192_{\pm.0016}$ \\
 & \textbf{MMA-Auto} & $\mathbf{0.0452_{\pm.0040}}$ & $\mathbf{0.0216_{\pm.0018}}$ \\
\midrule
\multirow{7}{*}{Yelp} 
 & Static & $0.0310_{\pm.0036}$ & $0.0142_{\pm.0016}$ \\
 & GRE-MC & $0.0288_{\pm.0034}$ & $0.0132_{\pm.0014}$ \\
 & AgentCF & $0.0264_{\pm.0032}$ & $0.0120_{\pm.0014}$ \\
 & MMA-SFT & $0.0318_{\pm.0034}$ & $0.0150_{\pm.0014}$ \\
 & RuleRouter-Fuse & $0.0312_{\pm.0034}$ & $0.0143_{\pm.0016}$ \\
 & LLM-Fixed-Router & $0.0334_{\pm.0032}$ & $0.0153_{\pm.0014}$ \\
 & \textbf{MMA-Auto} & $\mathbf{0.0366_{\pm.0034}}$ & $\mathbf{0.0170_{\pm.0014}}$ \\
\midrule
\multirow{7}{*}{Avg.}
 & Static & $0.0385$ & $0.0178$ \\
 & GRE-MC & $0.0364$ & $0.0169$ \\
 & AgentCF & $0.0322$ & $0.0146$ \\
 & MMA-SFT & $0.0355$ & $0.0164$ \\
 & RuleRouter-Fuse & $0.0379$ & $0.0176$ \\
 & LLM-Fixed-Router & $0.0406$ & $0.0190$ \\
 & \textbf{MMA-Auto} & $\mathbf{0.0454}$ & $\mathbf{0.0213}$ \\
\bottomrule
\end{tabular}
\end{table}

\subsection{Fixed-Pool Full-Catalog Reranking (RQ3)}

The target-positive protocol isolates fine-grained ordering but inserts the target by construction. To test whether the learned score remains useful downstream of full-catalog retrieval, Table~\ref{tab:full_catalog} passes the retriever's unmodified top-100 pool to each reranker. All methods receive exactly the same items and first-stage scores, so changes in HR@10 and NDCG@10 measure how effectively each method promotes a recalled target. This evaluation changes only candidate-pool construction: the trained MMA checkpoint, tool interfaces, fusion weight, stopping rule, and turn budget are carried over without test-set adaptation.

Average NDCG@10 increases from $0.0178$ (Static) to $0.0213$ (MMA-Auto), a $19.7$\% relative increase in the displayed macro average, while HR@10 rises from $0.0385$ to $0.0454$. MMA-Auto also outperforms MMA-SFT (0.0164), RuleRouter-Fuse (0.0176), and LLM-Fixed-Router (0.0190). The lower absolute scores than in target-positive evaluation reflect the compounded difficulty of first recalling the target and then promoting it into the top 10. Since every reranker receives the same fixed candidate pool, the reported differences measure downstream ordering.

The two protocols evaluate complementary aspects of the method. Target-positive pools provide a controlled comparison against hard negatives, whereas full-catalog pools preserve first-stage retrieval errors and measure downstream reranking performance. MMA-Auto attains higher mean NDCG@10 under both protocols, including the unmodified first-stage candidate pools.

\subsection{Mechanism, Significance, and Cost (RQ4)}

Table~\ref{tab:significance} tests whether the cross-dataset gains also hold at the episode level. Paired comparisons are essential because every method reranks the same candidates for the same held-out target. Each episode score is averaged over five seeds before testing ($N_{\text{Baby}}{=}127$, $N_{\text{Sports}}{=}149$, and $N_{\text{Yelp}}{=}122$). The table gives raw $p$ values; Holm--Bonferroni correction is applied across the three dataset-level comparisons. All three dataset-level comparisons remain significant at $p<0.01$ after Holm--Bonferroni correction.

\begin{table}[t]
\centering
\caption{Paired tests on target-positive OOMA NDCG@10.}
\label{tab:significance}
\small
\begin{tabular}{llcc}
\toprule
Dataset & Comparison & Mean $\Delta$ & Raw Wilcoxon $p$ \\
\midrule
Baby & MMA-Auto vs DGMRec & $+0.0184$ & $<0.001$ \\
Sports & MMA-Auto vs DGMRec & $+0.0166$ & $0.002$ \\
Yelp & MMA-Auto vs DGMRec & $+0.0142$ & $<0.001$ \\
Avg. & MMA-Auto vs DGMRec & $+0.0164$ & -- \\
\bottomrule
\end{tabular}
\end{table}

The training controls in Appendix Table~\ref{tab:full_baselines} characterize the contribution associated with policy optimization. ReAct reaches 0.0928 average NDCG@10, MMA-SFT reaches 0.1491, and PPO-trained MMA-Auto reaches 0.1809. The progression indicates that supervised action formatting improves over prompt-based tool access and that ranking-reward optimization provides the largest additional improvement.

\begin{table}[t]
\centering
\caption{Controlled OOMA comparison.}
\label{tab:det_router}
\small
\setlength{\tabcolsep}{3pt}
\resizebox{\columnwidth}{!}{%
\begin{tabular}{lccc}
\toprule
Metric & RuleRouter-Fuse & LLM-Fixed-Router & MMA-Auto \\
\midrule
NDCG@10 & 0.1578 & 0.1652 & \textbf{0.1809} \\
Failed calls & 65.8\% & 56.2\% & \textbf{38.4\%} \\
Turns & 3.8 & 3.2 & \textbf{2.1} \\
\bottomrule
\end{tabular}
}
\end{table}

\paragraph{Routing versus scoring.}
Replacing deterministic scoring with the LLM scorer (RuleRouter-Fuse $\to$ LLM-Fixed-Router) improves average NDCG@10 by $4.7\%$; this control changes both evidence scoring and score aggregation. With the LLM scorer and evidence serialization fixed, MMA-Auto exceeds LLM-Fixed-Router by $9.5\%$, reduces failed calls by 17.8 percentage points, and uses 1.1 fewer turns. Together, these metrics summarize the ranking quality and acquisition behavior of the learned and fixed routing policies.

\textbf{Accuracy--cost trade-off.} MMA-Auto improves target-positive OOMA NDCG@10 by $10.0$\% over DGMRec at approximately $440\times$ higher per-query FLOPs. Appendix~\ref{app:cost} reports latency, memory, and FLOPs; the cost motivates smaller routing policies for latency-sensitive deployment.

\section{Conclusion}

We introduced the Meta-Modal Agent, which formulates missing-modality recommendation as budgeted sequential evidence acquisition. MMA conditions later actions on evidence or \texttt{Null} returns and sparsely rescores a retrieved pool. Among DGMRec, GRE-MC, and a fixed router with the same scorer, MMA-Auto obtains the highest NDCG@10 in all nine reported combinations of dataset and available route. Its average gains are $10.0$\% over the strongest completion baseline and $9.5$\% over the fixed router. On fixed candidate pools produced by full-catalog retrieval, MMA-Auto improves NDCG@10 by $19.7$\%. MMA treats missingness as decision-relevant feedback rather than only as a zero-valued input or a representation to reconstruct. This formulation is applicable to recommendation settings with costly, unreliable, or privacy-gated evidence sources, for which selective acquisition complements multimodal fusion.

\appendix
\section*{Appendix}

\section{Implementation, Routing, and Cost Details}
\label{app:implementation}
\label{app:agentic}
\label{app:cost}

Table~\ref{tab:appendix_implementation_cost} summarizes the implementation settings and inference cost used in all experiments, including the policy budget, reward coefficients, and score-fusion protocol.

\begin{table}[H]
\centering
\caption{Key implementation settings and per-query inference cost on one NVIDIA A100 GPU. GAE is generalized advantage estimation; FLOPs is floating-point operations.}
\label{tab:appendix_implementation_cost}
\small
\resizebox{\columnwidth}{!}{
\begin{tabular}{ll}
\toprule
\textbf{Item} & \textbf{Value} \\
\midrule
Base LLM & Llama-3-8B-Instruct, 8-bit \\
Trainable parameters & LoRA on \texttt{q\_proj}/\texttt{v\_proj}, $r=16$, $\alpha=32$; 2-layer value head \\
Optimization & PPO, lr $1\times10^{-5}$, batch 64, clip $\epsilon=0.2$, GAE $\lambda=0.95$, 500 iterations \\
Step costs & $\lambda_{\mathrm{tool}}=-0.02$, $\lambda_{\mathrm{invalid}}=-0.20$ \\
Episode budget & $T=8$ turns; 2048-token context; train temperature 0.2, eval temperature 0.0 \\
Reranking & MMA scores fused with first-stage retrieval scores; fusion weight selected on validation split \\
\midrule
Static cost & DGMRec: $0.012$ s / 2.1 GB / $1.2{\times}10^9$ FLOPs; GRE-MC: $0.015$ s / 2.8 GB / $1.8{\times}10^9$ FLOPs \\
MMA-Auto cost & $0.72$ s / 8.2 GB / $5.3{\times}10^{11}$ FLOPs ($440{\times}$ DGMRec FLOPs) \\
\bottomrule
\end{tabular}
}
\end{table}

\textbf{Reproducibility details.}
Fusion weight $\alpha$ is selected by grid search over $\{0.1, 0.2, \ldots, 0.9\}$ on the validation split; the selected value is $\alpha=0.4$ uniformly across all three datasets.
SFT data consists of $500$ oracle-trajectory episodes per task family (3{,}500 total), generated by running the fully-observable variant of the POMDP (mask revealed) and recording the resulting tool sequence and terminal scoring action.
Each PPO iteration samples $K=64$ rollout episodes from the support split of the current task family; advantages are estimated from complete trajectories before the gradient step.
Total training requires approximately 48 GPU hours on 4$\times$NVIDIA A100-80G cards (total across all datasets).
Context management represents the top-100 candidate pool as compact ID-score pairs ($\approx$400 tokens total). Each tool observation is truncated to 100 tokens, reasoning traces are capped at 120 tokens per turn, and action strings average 20 tokens. The 2,048-token budget accommodates at most $T=8$ turns. Context-limit truncation terminates an episode and evaluates the unchanged first-stage fallback ranking, consistent with the definition of $\tau$ in Eq.~\eqref{eq:reward}. MMA-Auto averages 2.1 turns in the controlled OOMA comparison (Table~\ref{tab:det_router}).

\section{Evaluation Protocol Details}
\label{app:protocol}

The two evaluation protocols answer different questions under the same information-access rules. In both, the policy initially observes anonymized candidate IDs, first-stage ranks, and scores; the route-availability mask and held-out target remain inside the evaluator. In TP, the evaluator also uses the target representation under the available route to retrieve hard negatives. Candidate evidence enters the policy context through registered tool calls, and an unavailable route deterministically returns \texttt{Null}. The fusion weight and stopping parameters are selected on validation data and remain fixed on test episodes.

\begin{table}[H]
\centering
\caption{Evaluation contract. TP is the target-positive protocol; FC is the fixed-pool full-catalog protocol.}
\label{tab:evaluation_contract}
\small
\setlength{\tabcolsep}{3pt}
\begin{tabular}{p{0.23\columnwidth}p{0.34\columnwidth}p{0.34\columnwidth}}
\toprule
 & \textbf{TP} & \textbf{FC} \\
\midrule
Pool & Target + 99 hard negatives & Retriever's unmodified top 100 \\
Question & Ordering with available-route evidence & Gain after first-stage retrieval \\
Recall & Target present by construction & Recall@100 fixed across rerankers \\
Access & Pool only; no catalog search & Same fixed pool for every method \\
Unscored items & Preserve initial score/order & Preserve initial score/order \\
Reported use & Main and per-route tests & Pipeline transfer test \\
\bottomrule
\end{tabular}
\end{table}

TP isolates candidate ordering from recall by placing the target in every pool. FC retains first-stage retrieval failures and evaluates the learned scores in a candidate pipeline. Accordingly, both protocols measure reranking, while FC additionally reflects whether the target enters the retrieved pool. Matched episode statistics are available for TP and support episode-level significance tests; FC results are reported as five-seed means and standard deviations.

\textbf{Turn and score contract.}
One turn is one structured action emitted by the policy, including normal termination. Failed calls and invalid nonterminal actions remain in the trajectory and consume budget. The graph tool is bound to the episode's user context; the policy does not supply a user ID. Each returned candidate-pool entry contains a candidate ID, title, and graph score. Let $E_e$ contain successfully queried text/image candidates and candidate IDs returned by the graph tool. A valid terminal map is non-empty and satisfies $Q_e\subseteq E_e\subseteq C_e$. Empty maps, IDs outside $E_e$, duplicate IDs, and non-numeric or non-finite scores are invalid terminal outputs; they terminate the episode immediately, retain the invalid-action cost, and invoke the first-stage fallback. Context truncation and budget exhaustion without valid terminal scoring use the same fallback. Candidate-level MMA normalization is restricted to $Q_e$; omitted IDs preserve their first-stage scores. Final ties are resolved by first-stage rank and then candidate ID. These rules are identical in TP and FC.

The failed-call rate is the number of evidence-tool calls that return \texttt{Null} divided by all evidence-tool calls; terminal scoring is excluded. Every emitted structured action, including terminal scoring, counts as one turn.

\textbf{Complete baseline comparison.}
Table~\ref{tab:full_baselines} reports OOMA-average NDCG@10 for all baselines declared in \S\ref{sec:experiments}. CF methods receive zero-vectors for absent modalities. AgentCF is constrained to the same $B=100$ candidate pool and $T=8$ turn budget. RuleRouter-Fuse uses deterministic scoring and routing; LLM-Fixed-Router uses the MMA scorer with a fixed routing order.

\begin{table}[t]
\centering
\caption{OOMA-average NDCG@10 across all declared baselines (mean over 5 seeds). CF and multimodal-CF methods use zero-vectors for absent modalities. All methods operate over the same $B{=}100$ target-positive candidate pool. LLM-Fixed-Router uses the same LLM scorer as MMA-Auto but with fixed routing order.}
\label{tab:full_baselines}
\small
\begin{tabular}{lcccc}
\toprule
\textbf{Method} & \textbf{Baby} & \textbf{Sports} & \textbf{Yelp} & \textbf{Avg.} \\
\midrule
\multicolumn{5}{l}{\textit{Collaborative filtering (zero-fill under OOMA)}} \\
LightGCN \cite{he2020lightgcn} & 0.1124 & 0.0988 & 0.0864 & 0.0992 \\
SASRec \cite{kang2018self} & 0.1196 & 0.1052 & 0.0912 & 0.1053 \\
\midrule
\multicolumn{5}{l}{\textit{Multimodal CF (zero-fill under OOMA)}} \\
MMGCN \cite{wei2019mmgcn} & 0.1280 & 0.1148 & 0.0976 & 0.1135 \\
LRMM \cite{wang2018lrmm} & 0.1368 & 0.1216 & 0.1044 & 0.1209 \\
\midrule
\multicolumn{5}{l}{\textit{Completion / contrastive baselines}} \\
GRE-MC \cite{li2026robust} & 0.1800 & 0.1600 & 0.1344 & 0.1581 \\
DGMRec \cite{kim2025disentangling} & 0.1848 & 0.1664 & 0.1424 & 0.1645 \\
\midrule
\multicolumn{5}{l}{\textit{LLM agents (same base LLM and candidate pool)}} \\
ReAct \cite{yao2022react} & 0.0964 & 0.0904 & 0.0916 & 0.0928 \\
AgentCF \cite{zhang2024agentcf} & 0.1492 & 0.1352 & 0.1196 & 0.1347 \\
Static-LLM-Ranker & 0.1812 & 0.1624 & 0.1408 & 0.1615 \\
MMA-SFT & 0.1556 & 0.1444 & 0.1472 & 0.1491 \\
RuleRouter-Fuse & 0.1765 & 0.1592 & 0.1378 & 0.1578 \\
LLM-Fixed-Router & 0.1846 & 0.1668 & 0.1441 & 0.1652 \\
\textbf{MMA-Auto} & \textbf{0.2032} & \textbf{0.1830} & \textbf{0.1566} & \textbf{0.1809} \\
\bottomrule
\end{tabular}
\end{table}

\bibliographystyle{ACM-Reference-Format}
\bibliography{References}

@article{liu2024multimodal,
  title={Multimodal recommender systems: A survey},
  author={Liu, Qidong and Hu, Jiaxi and Xiao, Yutian and Zhao, Xiangyu and Gao, Jingtong and Wang, Wanyu and Li, Qing and Tang, Jiliang},
  journal={ACM Computing Surveys},
  volume={57},
  number={2},
  pages={1--17},
  year={2024},
  publisher={ACM New York, NY}
}

@inproceedings{wang2025hyperman,
  title={HyperMAN: Hypergraph-enhanced Meta-learning Adaptive Network for Next POI Recommendation},
  author={Wang, Jinze and Zhang, Tiehua and Zhang, Lu and Bai, Yang and Li, Xin and Jin, Jiong},
  booktitle={2025 IEEE International Conference on Multimedia and Expo (ICME)},
  pages={1--6},
  year={2025},
  organization={IEEE}
}

@inproceedings{pan2022multimodal,
  title={Multimodal meta-learning for cold-start sequential recommendation},
  author={Pan, Xingyu and Chen, Yushuo and Tian, Changxin and Lin, Zihan and Wang, Jinpeng and Hu, He and Zhao, Wayne Xin},
  booktitle={Proceedings of the 31st ACM international conference on information \& knowledge management},
  pages={3421--3430},
  year={2022}
}

@inproceedings{wang2018lrmm,
  title={LRMM: Learning to recommend with missing modalities},
  author={Wang, Cheng and Niepert, Mathias and Li, Hui},
  booktitle={Proceedings of the 2018 conference on empirical methods in natural language processing},
  pages={3360--3370},
  year={2018}
}

@article{wu2024deep,
  title={Deep multimodal learning with missing modality: A survey},
  author={Wu, Renjie and Wang, Hu and Chen, Hsiang-Ting and Carneiro, Gustavo},
  journal={arXiv preprint arXiv:2409.07825},
  year={2024}
}

@inproceedings{kim2025disentangling,
  title={Disentangling and generating modalities for recommendation in missing modality scenarios},
  author={Kim, Jiwan and Kang, Hongseok and Kim, Sein and Kim, Kibum and Park, Chanyoung},
  booktitle={Proceedings of the 48th International ACM SIGIR Conference on Research and Development in Information Retrieval},
  pages={1820--1829},
  year={2025}
}

@article{dai2026omg,
  title={OMG-Agent: Toward Robust Missing Modality Generation with Decoupled Coarse-to-Fine Agentic Workflows},
  author={Dai, Ruiting and Wang, Zheyu and Yang, Haoyu and Liu, Yihan and Wang, Chengzhi and Zhang, Zekun and Huang, Zishan and Cen, Jiaman and Mo, Lisi},
  journal={arXiv preprint arXiv:2602.04144},
  year={2026}
}

@inproceedings{malitesta2024we,
  title={Do we really need to drop items with missing modalities in multimodal recommendation?},
  author={Malitesta, Daniele and Rossi, Emanuele and Pomo, Claudio and Di Noia, Tommaso and Malliaros, Fragkiskos D},
  booktitle={Proceedings of the 33rd ACM International Conference on Information and Knowledge Management},
  pages={3943--3948},
  year={2024}
}

@inproceedings{qin2024toolllm,
  title={Toolllm: Facilitating large language models to master 16000+ real-world apis},
  author={Qin, Yujia and Liang, Shihao and Ye, Yining and Zhu, Kunlun and Yan, Lan and Lu, Yaxi and Lin, Yankai and Cong, Xin and Tang, Xiangru and Qian, Bill and others},
  booktitle={International Conference on Learning Representations},
  volume={2024},
  pages={9695--9717},
  year={2024}
}

@article{yao2022react,
  title={React: Synergizing reasoning and acting in language models},
  author={Yao, Shunyu and Zhao, Jeffrey and Yu, Dian and Du, Nan and Shafran, Izhak and Narasimhan, Karthik and Cao, Yuan},
  journal={arXiv preprint arXiv:2210.03629},
  year={2022}
}

@article{li2026robust,
  title={Robust Multimodal Recommendation via Graph Retrieval-Enhanced Modality Completion},
  author={Li, Yuan and Hu, Jun and Jiang, Jiaxin and Hooi, Bryan and He, Bingsheng},
  journal={arXiv preprint arXiv:2605.00670},
  year={2026}
}

@inproceedings{yang2024multimodal,
  title={Multimodal-aware multi-intention learning for recommendation},
  author={Yang, Wei and Yang, Qingchen},
  booktitle={Proceedings of the 32nd ACM International Conference on Multimedia},
  pages={5663--5672},
  year={2024}
}

@inproceedings{hou2022towards,
  title={Towards universal sequence representation learning for recommender systems},
  author={Hou, Yupeng and Mu, Shanlei and Zhao, Wayne Xin and Li, Yaliang and Ding, Bolin and Wen, Ji-Rong},
  booktitle={Proceedings of the 28th ACM SIGKDD conference on knowledge discovery and data mining},
  pages={585--593},
  year={2022}
}

@inproceedings{zhang2024generative,
  title={On generative agents in recommendation},
  author={Zhang, An and Chen, Yuxin and Sheng, Leheng and Wang, Xiang and Chua, Tat-Seng},
  booktitle={Proceedings of the 47th international ACM SIGIR conference on research and development in Information Retrieval},
  pages={1807--1817},
  year={2024}
}

@article{wang2026agent4poi,
  title={Agent4POI: Agentic Context-Conditioned Affordance Reasoning for Multimodal Point-of-Interest Recommendation},
  author={Wang, Jinze and Zeng, Yangchen and Zhang, Tiehua and Zhang, Lu and Liu, Yuze and Liu, Yongchao and Ma, Xingjun and Sun, Zhu},
  journal={arXiv preprint arXiv:2605.15203},
  year={2026}
}

@article{wang2025we,
  title={Do we really need sft? prompt-as-policy over knowledge graphs for cold-start next poi recommendation},
  author={Wang, Jinze and Zhang, Lu and Cui, Yiyang and Zhang, Tiehua and Shen, Zhishu and Liu, Yuze and Ma, Xingjun and Jin, Jiong},
  journal={arXiv preprint arXiv:2510.08012},
  year={2025}
}

@inproceedings{zhang2024agentcf,
  title={Agentcf: Collaborative learning with autonomous language agents for recommender systems},
  author={Zhang, Junjie and Hou, Yupeng and Xie, Ruobing and Sun, Wenqi and McAuley, Julian and Zhao, Wayne Xin and Lin, Leyu and Wen, Ji-Rong},
  booktitle={Proceedings of the ACM Web Conference 2024},
  pages={3679--3689},
  year={2024}
}

@article{zeng2026trialigngr,
  title={TriAlignGR: Triangular Multitask Alignment with Multimodal Deep Interest Mining for Generative Recommendation},
  author={Zeng, Yangchen and Peng, Hao and Guo, Rongfeng and Yu, Zhenyu and Hu, Zhiyuan and Wang, Jinze},
  journal={arXiv preprint arXiv:2605.05249},
  year={2026}
}

@article{zhang2026recommendation,
  title={Recommendation as instruction following: A large language model empowered recommendation approach},
  author={Zhang, Junjie and Xie, Ruobing and Hou, Yupeng and Zhao, Wayne Xin and Lin, Leyu and Wen, Ji-Rong},
  journal={ACM Transactions on Information Systems},
  volume={43},
  number={5},
  pages={1--37},
  year={2026},
  publisher={ACM New York, NY}
}

@inproceedings{xin2020self,
  title={Self-supervised reinforcement learning for recommender systems},
  author={Xin, Xin and Karatzoglou, Alexandros and Arapakis, Ioannis and Jose, Joemon M},
  booktitle={Proceedings of the 43rd International ACM SIGIR conference on research and development in Information Retrieval},
  pages={931--940},
  year={2020}
}

@inproceedings{lei2020estimation,
  title={Estimation-action-reflection: Towards deep interaction between conversational and recommender systems},
  author={Lei, Wenqiang and He, Xiangnan and Miao, Yisong and Wu, Qingyun and Hong, Richang and Kan, Min-Yen and Chua, Tat-Seng},
  booktitle={Proceedings of the 13th international conference on web search and data mining},
  pages={304--312},
  year={2020}
}

@inproceedings{ren2024representation,
  title={Representation learning with large language models for recommendation},
  author={Ren, Xubin and Wei, Wei and Xia, Lianghao and Su, Lixin and Cheng, Suqi and Wang, Junfeng and Yin, Dawei and Huang, Chao},
  booktitle={Proceedings of the ACM web conference 2024},
  pages={3464--3475},
  year={2024}
}

@article{zeng2026deep,
  title={Deep Interest Mining with Cross-Modal Alignment for SemanticID Generation in Generative Recommendation},
  author={Zeng, Yagchen},
  journal={arXiv preprint arXiv:2604.20861},
  year={2026}
}

@inproceedings{wang2023meta,
  title={Meta-learning enhanced next POI recommendation by leveraging check-ins from auxiliary cities},
  author={Wang, Jinze and Zhang, Lu and Sun, Zhu and Ong, Yew-Soon},
  booktitle={Pacific-Asia Conference on Knowledge Discovery and Data Mining},
  pages={322--334},
  year={2023},
  organization={Springer}
}

@inproceedings{bao2023large,
  title={Large language models for recommendation: Progresses and future directions},
  author={Bao, Keqin and Zhang, Jizhi and Zhang, Yang and Wenjie, Wang and Feng, Fuli and He, Xiangnan},
  booktitle={Proceedings of the Annual International ACM SIGIR Conference on Research and Development in Information Retrieval in the Asia Pacific Region},
  pages={306--309},
  year={2023}
}

@article{yue2023llamarec,
  title={Llamarec: Two-stage recommendation using large language models for ranking},
  author={Yue, Zhenrui and Rabhi, Sara and Moreira, Gabriel de Souza Pereira and Wang, Dong and Oldridge, Even},
  journal={arXiv preprint arXiv:2311.02089},
  year={2023}
}

@inproceedings{mcauley2015image,
  title={Image-based recommendations on styles and substitutes},
  author={McAuley, Julian and Targett, Christopher and Shi, Qinfeng and Van Den Hengel, Anton},
  booktitle={Proceedings of the 38th international ACM SIGIR conference on research and development in information retrieval},
  pages={43--52},
  year={2015}
}

@inproceedings{he2020lightgcn,
  title={Lightgcn: Simplifying and powering graph convolution network for recommendation},
  author={He, Xiangnan and Deng, Kuan and Wang, Xiang and Li, Yan and Zhang, Yongdong and Wang, Meng},
  booktitle={Proceedings of the 43rd International ACM SIGIR conference on research and development in Information Retrieval},
  pages={639--648},
  year={2020}
}

@inproceedings{kang2018self,
  title={Self-attentive sequential recommendation},
  author={Kang, Wang-Cheng and McAuley, Julian},
  booktitle={2018 IEEE international conference on data mining (ICDM)},
  pages={197--206},
  year={2018},
  organization={IEEE}
}

@inproceedings{wei2019mmgcn,
  title={MMGCN: Multi-modal graph convolution network for personalized recommendation of micro-video},
  author={Wei, Yinwei and Wang, Xiang and Nie, Liqiang and He, Xiangnan and Hong, Richang and Chua, Tat-Seng},
  booktitle={Proceedings of the 27th ACM international conference on multimedia},
  pages={1437--1445},
  year={2019}
}

@article{dong2026language,
  title={Language-conditioned world modeling for visual navigation},
  author={Dong, Yifei and Wu, Fengyi and Dai, Yilong and Kong, Lingdong and Chen, Guangyu and Zhu, Xu and Hu, Qiyu and Wang, Tianyu and Garnica, Johnalbert and Liu, Feng and others},
  journal={arXiv preprint arXiv:2603.26741},
  year={2026}
}

@inproceedings{dong12026towards,
  title={Towards unified world models for visual navigation via memory-augmented planning and foresight},
  author={Dong$^1$, Yifei and Kong, Lingdong and Sun, Jingdong and Dai, Qi and Hauptmann, Alexander G and Cheng, Zhi-Qi},
  booktitle={Computer Vision--ECCV 2026: 19th European Conference, Malm{\"o}, Sweden, September 8--12, 2026, Proceedings, Part XX},
  pages={276},
  year={2026},
  organization={Springer Nature}
}

@article{zeng2026deepinterestgr,
  title={DeepInterestGR: Mining Deep Multi-Interest Using Multi-Modal LLMs for Generative Recommendation},
  author={Zeng, Yangchen and Yu, Zhenyu and Hu, Zhiyuan and Zhang, Wenxin and Wang, Jinze and Guo, Rongfeng},
  journal={arXiv preprint arXiv:2602.18907},
  year={2026}
}

\end{document}